\documentclass[aps,rmp,twocolumn,amsmath,amssymb,graphicx]{revtex4}

\usepackage{epsfig}
\usepackage{epstopdf}
\usepackage{euscript}
\usepackage{amsfonts}
\usepackage{float}
\usepackage{url}

\def\bra#1{{\langle#1|}}

\def\cg(#1,#2)(#3,#4)(#5,#6){\bra{#1,#2,#3,#4}#5,#6\rangle}

\def\threej(#1,#2)(#3,#4)(#5,#6){\begin{pmatrix}#1&#3&#5\\#2&#4&#6\end{pmatrix}}
\def\sixj(#1,#2,#3)(#4,#5,#6){\begin{Bmatrix}#1&#2&#3\\#4&#5&#6\end{Bmatrix}}
\def\ninej(#1,#2,#3)(#4,#5,#6)(#7,#8,#9){\begin{Bmatrix}#1&#2&#3\\#4&#5&#6\\#7&#8&#9\end{Bmatrix}}

\newlength{\defbaselineskip}
\newcommand{\setlinespacing}[1]%
           {\setlength{\baselineskip}{#1 \defbaselineskip}}

\begin{document}

\title{Rules for collaborative scientific writing} 

\date{\today}

\author{D.~Budker}
\affiliation{Helmholtz Institute Mainz, Johannes Gutenberg University, 55099 Mainz, Germany}
\affiliation{Department of Physics, University of California at Berkeley, Berkeley, California 94720-7300, USA}
\affiliation{Nuclear Science Division, Lawrence Berkeley National Laboratory, Berkeley, California 94720, USA}

\author{Derek F. Jackson Kimball}
\affiliation{Department of Physics, California State University -- East Bay, Hayward, California 94542-3084, USA}

\maketitle

We have recently published a brief guide concerning collaborative scientific writing \cite{Bud16}. Here we present the original manuscript upon which this recent guide was based:

\section{Introduction}

\begin{figure}
\center
\includegraphics[width=3.25 in]{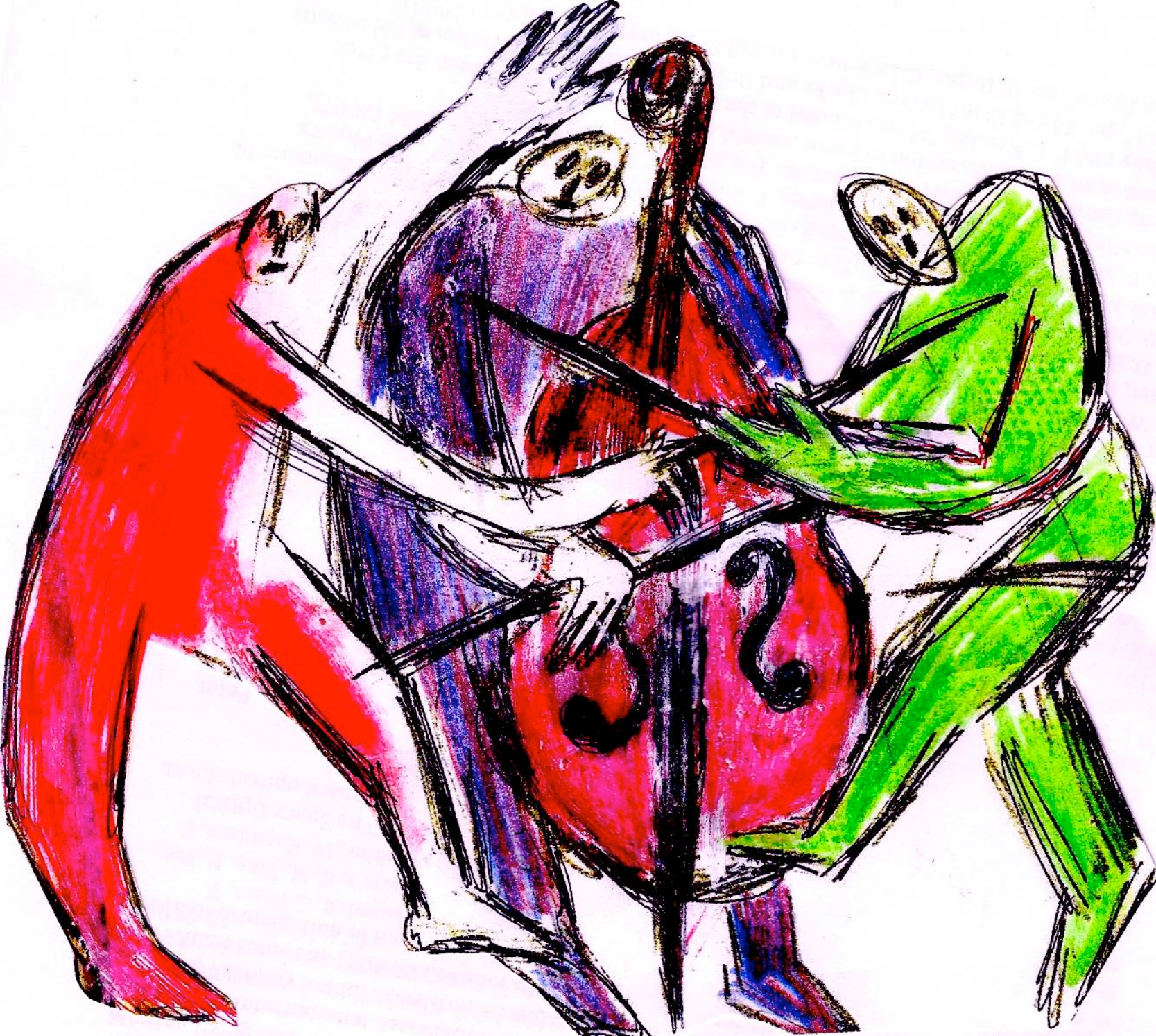}
\caption{How can many musicians collaborate to make beautiful music? Drawing courtesy of Olga Budker.}
\label{Fig:musicians}
\end{figure}

Several years ago, one of us, having noticed that inexperienced scientists tend to make largely the same mistakes while writing their first papers, was compelled to write a one-page note \cite{Bud06} summarizing some dos and don'ts intended to help take care of common problems before they occur. Since these days the majority of research papers are written collaboratively by groups of co-authors (Fig.~\ref{Fig:musicians}), we are compelled to extend the recommendations of \citet{Bud06} to collaborative writing as we observe groups of co-authors falling into the same traps again and again.

\section{Composing author}

The paper should have a single {\it{composing}} or {\it{lead author}}. This person maintains the {\it{master copy}} of the manuscript and incorporates input from all co-authors. Ideally, the composing author should remain the same from the conception of the manuscript to its publication. Of course, life is complicated, and it may not always work out this way. Upon agreement among the co-authors, the composing author can change, but it is imperative for the manuscript to be ``owned'' by no more that one person at any given time. In our experience, if a group of scientists has agreed to write a paper together, it is not hard to find a volunteer to be the composing author. Ideally, the composing author should be the one who has done most of the work and is most familiar with the details of the project, giving the composing author a deep and broad perspective. Thinking carefully at the beginning of the writing project about who shall lead and guide it also forces everyone involved to think the project through before starting it. Whereas the lead author is assigned a special role, she or he must not forget the ever-so-wise words of Voltaire (and Peter Parker's uncle): ``With great power comes great responsibility.''

\section{Co-authors}

Who should be a co-author of the paper? Two key questions we ask are: ``did the researcher make a meaningful contribution to the project?'' and ``does the researcher understand the complete work well enough to explain and defend it to colleagues?''

Becoming a co-author of a paper is a serious commitment: that person's scientific reputation is now wedded to the validity of the paper. We have found that it makes sense to be reasonably inclusive: generally, people who feel that they do not meet the threshold for co-authorship will excuse themselves from the author list. It is much more damaging to the collaboration to leave someone off the author list who feels that they should be included than it is to include someone who has made marginal contributions. The omission can create hard feelings and possibly violate scientific ethics.

In some cases, it can work well if the first drafts of different manuscript sections are composed by different team members in accordance with their strengths: a theorist could write a section on calculations and an experimentalist could write a section on measurements, for example. If your collaboration uses this model, the composing author should be the one to incorporate the different sections into a coherent story.

\section{Start writing early}

We always urge our lab groups and collaborators to start writing before the research project is complete. A paper usually begins with an explanation of the motivation for the project and a survey of related work. What better time to create this section than before you have fully plunged into the study?
Also, doing this bit at an earlier stage helps to delineate the paper's logical path and to provide a guide for the research ahead. When the survey of earlier work is prepared in advance, it is easier to foresee, for instance, what might need to be measured to choose between possible alternative interpretations of the data.

It is also useful to create an overall outline for the paper early on. The composing author should write the outline, discuss it with the team, and the team should come to a consensus.

Such a framework helps to crystallize the goals of the project, forces the group to address key scientific questions from the outset and keeps the work focused. More than once, we have completed a study and reconfigured the experimental apparatus for some new investigation, only to discover while writing the paper
that we had overlooked a detail that requires more data. The outline can help to avert this misstep.

An old saying we have is: ``six months in the laboratory saves an hour in the library!''

\section{Only the composing author should edit the master copy}

Authors other than the composing author should {\textbf{NOT}} edit the master copy himself or herself. They should send suggestions, corrections, and new pieces of text to the composing author, and the composing author should incorporate these pieces into the master document. When the co-authors wish to propose minor edits, they should send the composing author the suggestions either in a list (for example, in the text of an e-mail), or in the form of markup in the text.

\section{Carefully consider all input}

When the composing author receives input from a co-author, she or he should critically examine the input; in case the composing author agrees, she or he should incorporate the input into the master. If she or he does not agree or does not understand the co-author's suggestions (which happens frequently), these suggestions should be brought to a discussion with the co-author or with a bigger group of co-authors.

\section{Do not ignore suggestions}

No suggestions by co-authors should be ignored, although this does not mean that they should all be incorporated. A discussion where a collective decision is made is the way to go. There are few things more frustrating to a co-author than reading a new draft of the manuscript and once again seeing the very same problem that the co-author had previously pointed out (and that was not subsequently discussed)! Ideally, the suggestions of co-authors should be treated like the suggestions of referees: taken very seriously and either agreed to or responded to in a comprehensive manner.

\section{Do not disappear}

Scientists are people of many interests, and at times one might encounter co-authors ``disappearing into a black hole.'' This is frustrating, especially for the composing author. We recommend defining a {\it{maximum response time}} (reasonably something like a few days) by when at least an acknowledgement of receipt of a communication should be sent. The composing author is not allowed anywhere near a black hole for the course of the entire writing project.

\section{Order of the author list}

Order of authors is often a source of contention, especially since different communities have different unwritten rules. The high-energy community tends to adhere to a strict alphabetical order of authors, while, for instance, in Atomic, Molecular, and Optical (AMO) physics, the students and post-docs central to the project usually go first, while the principal investigator is last. But what do you do if there is more than one student or a student and a post-doc who contributed equally? And what do you do if the work is a collaboration among several laboratories -- which laboratory goes first? Should the authors be grouped by the laboratory, or should there be a common ``mixed'' list?

We offer the following advice. First of all, think about the author order from the start. Moreover, nowadays many journals offer or demand authors-contribution sections, and these should be used -- and used wisely -- even in those cases where the journal editors do not require it. From the authors' contribution section it should become clear who did what in terms of the work and also who is primarily responsible for which parts of the manuscript (for example, sections on theory and experiment). All else equal, and in deference to the culture of the particular discipline, it is best if the composing author is also the first author. Ultimately, co-authors should not worry about this too much. We have witnessed heated arguments over who should be the first author; but 10 or 20 years later, we can see that it did not really matter that much in the end.

\section{Tension}

If there is a tension among the co-authors, for instance, on the order as discussed above, it should fall upon the Principal Investigator (PI) to figure it out (this is what the PI is paid the ``big bucks'' for, after all). Sometimes consensus just cannot be reached: as the senior person responsible for the project and the group, ultimately it is the PI's responsibility to do what is best and make the tough decision. As in every important decision about the manuscript, a detailed reasoning for the ``verdict'' should be provided for transparency. In the case of multiple PIs, they will just have to work it out like adults.

\section{Read the manuscript}

\noindent Repeating a point highlighted by \citet{Bud06}: 

~\

\noindent ``This one is a must: read the finished manuscript!''

~\

\noindent This holds for every co-author.

\section{Consensus for submission}

Each co-author needs to explicitly agree that the paper is ready for journal submission or posting to an e-print archive. Not only is this a courtesy to authors, but also is demanded by scientific ethics, not to mention the journal policy and legal considerations. All co-authors need to be informed of major events in the life of the manuscript, such as submission, resubmission, acceptance/rejection, receipt of the reviewers' comments, etc.

\section{Conclusions}

We emphasize that these rules are not arbitrary; they originate from frustrating and stressful experiences that we have lived through and witnessed. We hope you find these guidelines useful!

\acknowledgements

We are grateful to all our co-authors who helped us crystalize these rules, with special thanks to Prof. Surjeet Rajendran. We also thank Andreas Trabesinger for generously sharing his wisdom and advice.

\bibliography{collaborative-writing-ArXiv-version}

\end{document}